\documentclass[12pt]{article}

\usepackage[english]{babel}
\usepackage[usenames]{color}
\usepackage[cp1250]{inputenc}
\usepackage{amsfonts}
\usepackage{amsthm}
\usepackage{graphicx}
\usepackage{epsfig}

\theoremstyle{plain}

\begin{document}
\newcommand{\bea}{\begin{eqnarray}}
\newcommand{\eea}{\end{eqnarray}}
\newcommand{\be}{\begin{equation}}
\newcommand{\ee}{\end{equation}}
\newcommand{\beas}{\begin{eqnarray*}}
\newcommand{\eeas}{\end{eqnarray*}}
\newcommand{\bs}{\backslash}
\newcommand{\bc}{\begin{center}}
\newcommand{\ec}{\end{center}}

\title{Combinatorial invariants for\\ graph isomorphism problem.}

\author{Jarek Duda}

\date{\it \footnotesize Jagiellonian University, Reymonta 4, 30-059 Kraków, Poland, \\
\textit{email:} dudaj@interia.pl}

\maketitle

\begin{abstract}
Presented approach in polynomial time calculates large number of
invariants for each vertex, which won't change with graph
isomorphism and should fully determine the graph. For example
numbers of closed paths of length $k$ for given starting vertex,
what can be though as the diagonal terms of $k$-th power of the
adjacency matrix. For $k=2$ we would get degree of verities
invariant, higher describes local topology deeper. Now if two graphs
are isomorphic, they have the same set of such vectors of invariants
- we can sort theses vectors lexicographically and compare them. If
they agree, permutations from sorting allow to reconstruct the
isomorphism. I'm presenting arguments that these invariants should
fully determine the graph, but unfortunately I can't prove it in
this moment. This approach can give hope, that maybe P=NP - instead
of checking all instances, we should make arithmetics on these large
numbers.
\end{abstract}
\section{Introduction}
 We have some undirected graphs, given by the adjacency
matrix
$$G_1,G_2\in \{0,1\}^{n\times n}$$
We would like to check if there is a permutation matrix:
$$P\in \{0,1\}^{n\times n}:\ P^T P=\mathbf{1}_n,\ G_1=P G_2 P^T$$
So we have to check if $G_1,\ G_2$ are similar and if the
similarity matrix is permutation.\\
The similarity matrix could be found using numerical methods, which
are asymptotic - the problem is to estimate when to be sure if we
won't get permutation.\\

To check if the matrixes are similar, we can compare their
characteristic polynomials, what can be done in polynomial time.\\
If not - the graphs are not isomorphic, but if yes - we still don't
know if the similarity matrix is permutation, but it seems unlikely
that similarity matrix between two $\{0,1\}$ matrixes isn't
permutation.\\

\section{Algorithm} To get safer algorithm we will focus on, we can
compare diagonals of powers of the adjacency matrixes.\\

There is known and easy to check combinatorial property, that:
$$(G^k)_{ij}=\textrm{number of paths from }i\textrm{ to }j\textrm{ of
length }k$$ where in path edges and vertices can be repeated.\\

Without loss of generalities, we can assume that graph is connected,
so from Frobenius-Perron theorem it has unique dominant eigenvalue
($\leq n$) and corresponding eigenvector is nonnegative - the
diagonal terms of powers of the adjacency matrixes will increase
exponentially (length of numbers grows linearly) and in the limit has distribution as the eigenvector.\\

If the graphs are isomorphic, diagonals of above powers has to be
the same up to permutation. This time the isomorphism is suggested
by large numbers on the diagonal - we can just sort them.\\
Sometimes different vertices can give the same invariants - it
suggests some symmetry in the graph. In this case we have to be
careful if we would like to reconstruct the isomorphism - we should
build it neighbor by neighbor.\\

For second power the invariants ensure that degrees of vertices
agrees. \\
Higher powers checks local topology of vertex deeper and
deeper.\\
The length of numbers in powers of matrixes grows linearly with the
power, so calculating powers can be done in polynomial time.\\

The algorithm:\\
 \verb" For graph" $G$\verb":"\\
 \verb" For" $i,k=1,..,n$ \verb" calculate" $d_{ki}=(G^k)_{ii}$\\
 \verb" sort vectors "$\{(d_{ki})_k\}_i$ \verb" lexicographically"\\
gives us $n^2$ invariants in polynomial time.\\

If graphs are isomorphic, above invariants has to agree.\\
But if they agree, are graphs isomorphic? Do they determine graph uniquely? \\
I'll show that in 'generic' case it's true - we can even reconstruct the matrix.\\
Unfortunately I cannot prove in this moment that there are no graphs
that starting from above invariants, then eventually using some
standard techniques, we couldn't determine in polynomial time if
they are isomorphic, but it looks highly unprovable.\\

In practice we can make arithmetics modulo some large number and
just check a few steps $G\to G^2$ and try to reconstruct isomorphism
power by power. \\If something's wrong, we should
see it early, if not we should quickly get isomorphism to check.\\

\section{Reconstruction} We would like to have some nice
combinatorial procedure to
uniquely reconstruct the graph. Unfortunately I couldn't find it.\\
I will show algebraic construction, which is rather unpractical, but
should give unique graph in practically all - 'generic' cases.\\

Generally - there is some symmetric matrix $A\in \mathbb{R}^{n\times
n}$, but we only know
$$d_{ki}:=(A^k)_{ii} \quad \textrm{ for }\ i,k=1,..,n.$$
The matrix is real, symmetric so we can diagonalize it - there
exists $V,D$:
$$A=VDV^T\ \textrm{ where } V^T V=VV^T=\textbf{1}_n,\
D_{ij}=\lambda_i\delta_{ij}$$ where matrix $V$ is made of
eigenvectors: $Av_i=\lambda_i v_i$.\\

Observe that
$$\sum_i d_{ki}=Tr A^k=\sum_i \lambda_i^k$$
so we can reconstruct the characteristic polynomials determining the
spectrum.\\

We can present canonical base in the base of eigenvectors:
$e_i=\sum_j W_{ij} v_j$.\\
Writing this relation in columns we get: $\mathbf{1}_n=WV$, so
$W=V^T$,
$$e_i=\sum_j V_{ji} v_j$$
Finally we have (for $k=0$ we have $A^0=\mathbf{1}_n$):
$$d_{ki}=e_i^TA^ke_i=\left(\sum_j V_{ji} v_j^T\right)\left(\sum_l \lambda_l^k
V_{li} v_l\right)=\sum_j \lambda_j^k (V_{ji})^2 $$ We already know
the eigenvalues - for each $i$ we get interpolation problem. \\
Assume that there are no two equal eigenvalues - it's one of generic
property we would need. \\In this case, because the Vandermonde
matrix is reversible, we can find all $(V_{ij})^2$. We also see that
checking more than $n$ powers doesn't bring any new information.\\
If some eigenvalues repeats, we would find smaller number of
coefficient and have freedom to distribute our squares of
terms between them, what would complicate the next step.\\

We see that we have another problem - determine signs for nonzero
terms. \\Remember that $V$ is orthogonal:
$$\forall_{ij}\quad \sum_k V_{ik}V_{jk}=\delta_{ij}$$
and that in fact we are interested only in $A$:
$$\sum_k V_{ik}V_{jk}\lambda_k=A_{ij}$$
We see that multiplying whole column by $-1$ doesn't change $A$ - we
can fix signs in the first row of $V$ as we want.\\
Now using above two equations, and assuming that $A_{ij}\in \{0,1\}$
and there are no zero rows/columns in $A$ (graph is connected), we
have to determine the rest of signs. It can be thought as choosing
sings for 2-dimensional vectors, so that they sum up to one of two
points in $\mathbb{R}^2$. We know that there is one such assignment.
Coordinates of these vectors are some real numbers - in generic case
there shouldn't be second one.\\

So in 'generic' case - that there are no two the same eigenvalues
and that the signs can be assigned in unique way, the
invariants determine the graph up to isomorphism.\\

I've also found some relations between $d_{ki}$ and characteristic
polynomials of the matrix with removed column and row of the same
number - some kind of generalized Newton's identities. They could be
helpful for reconstructing the matrix.\\
The derivation is practically exactly Dan Kalman's derivation of
Newton's identities \cite{kal}, so I'm presenting it shortly and
referring
to the paper for details.\\
Denote $X:=x\mathbf{1}_n$,\\
$p(x)=x^n+a_{n-1}x^{n-1}+...+a_0=\det(X-A)$ - the characteristic
polynomial of $A$,\\$p_i(x)$ - characteristic polynomials of $A$ with removed $i$-th row and column.\\
From Ceyley-Hamiltonian theorem, we know that $p(A)=0$. Dividing it
by $X-A$, we will get:
$$(X-A)^{-1}p(X)=X^{n-1}+(A+a_{n-1}\textbf{1}_n)X^{n-2}+...
+(A^{n-1}+a_{n-1}A^{n-2}+...+a_1\textbf{1}_n)\textbf{1}_n$$

In this moment in \cite{kal} is taken trace of both side, the left
occurs to be $p'(x)$ - we get Newton's identities.\\
We can also be more subtle - take for example diagonal elements
($i$-th) - using formula for inverse matrix, we get $p_i$ on the
left side, so:
$$\sum_i p_i(x)=p'(x)$$
Using the right side, we get: \beas
p_i(x)=x^{n-1}+(A_{ii}+a_{n-1})x^{n-2}+((A^2)_{ii}+a_{n-1}A_{ii})x^{n-3}+...\\
...+((A^{n-1})_{ii}+a_{n-1}(A^{n-2})_{ii}+...+a_1)\eeas
 Finally having
$(d_{ki})_{k,i=1..n}$, summing over $i$ and using Newton's
identities we can find the characteristic polynomial and using above
relations find $(p_i)_{i=1..n}$ and spectrum of $A$ with removed
$i$-th row
and column. \\
On the other hand having $(p_i)_{i=1..n}$ we can sum them to get
$a_1,...,a_{n-1}$ and finally $d_{ki}$ using above identities. We
don't get the determinant ($a_0$) this way.

\section{More invariants}
In this moment we have $n$ independent invariants for each vertex,
which usually should determine the graph and needed $n-1$
multiplications of large matrixes with large numbers.\\
We see that if we would need less powers, the algorithm would be
much faster. We should achieve it using more invariants.\\
Unfortunately I still didn't prove fully determining of graph, but
using more invariants makes it even more probable.\\

In previous sections, for vertex $i$ from $N_{ki}=((A^k)_{ij})_j$ we
took only the diagonal term, because the rest permute in not known way.\\
We see that we could take the rest of terms of $N_{ki}$, but we
should forget about their order, but we should remember that for
different $k$, $j$ denotes the same vertex. \\For example for each
vertex $i$ we should sort lexicographically:
$$p_i:=\{\left(A_{ij},(A^2)_{ij},..,(A^n)_{ij}\right):j=1,..,n,j\neq i\}$$
For every power of $A$ we get this way $n(n-1)$ invariants. \\
Using the method from the end of previous section, we see that these
invariants are equivalent knowing for each $i$ list of
characteristic polynomials of $A$ with removed $i$-th row and one
column.\\
It suggest that again there is no point in using more than $n$
powers.\\

The other way of constructing invariants is using not only number of
pathes from given vertex, but also invariants of it's neighbors.
There is huge number of possibilities now - for example take sum of
some invariants of every neighbor of the vertex.\\

There have to be plenty of relations between these invariants. \\
We should now choose some invariants which fully determine the graph
and uses as small powers as possible.\\

To summarize - I didn't excluded cases that two graphs has the same
invariants and they are not isomorphic, but it looks extremely
improbable and probably it should be corrected in polynomial time by
trying to reconstruct the isomorphism using orders from sorting.

\end{document}